\documentclass[9pt,conference]{IEEEtran}
\usepackage{makeidx,epsfig}
\usepackage{graphicx}
\usepackage{setspace,graphicx}
\usepackage{booktabs}
\usepackage{multirow}
\usepackage{color,soul}
\usepackage{subfigure}

\usepackage[normalem]{ulem}
\usepackage{soul,xcolor}
\setstcolor{red}

\newcommand{\subparagraph}\textit{{}}

\newcommand{\Out}[1]{}

\usepackage{lipsum}

\makeatletter
\newcommand*\titleheader[1]{\gdef\@titleheader{#1}}
\AtBeginDocument{%
  \let\st@red@title\@title
  \def\@title{%
    \bgroup\normalfont\large\centering\@titleheader\par\egroup
    \vskip0em\st@red@title}
}
\titleheader{2019 IEEE International Conference on Blockchain (Blockchain)}
\makeatother

\title{ChainSplitter: Towards Blockchain-based Industrial IoT Architecture for Supporting Hierarchical Storage}

\begin{document}

\author{\IEEEauthorblockN{Gang Wang\IEEEauthorrefmark{1}, Zhijie Jerry Shi\IEEEauthorrefmark{1}, Mark Nixon\IEEEauthorrefmark{2}, Song Han\IEEEauthorrefmark{1}}
\IEEEauthorblockA{\IEEEauthorrefmark{1}University of Connecticut  \\
\IEEEauthorrefmark{2}Emerson Automation Solutions
\\ Email: \{gang.wang, zshi\}@uconn.edu, \ mark.nixon@emerson.com, \ song.han@uconn.edu}
}

\maketitle

\begin{abstract}
The fast developing Industrial Internet of Things (IIoT) technologies provide a promising opportunity to build large-scale systems to connect numerous heterogeneous devices into the Internet. Most existing IIoT infrastructures are based on a centralized architecture, which is easier for management but cannot effectively support immutable and verifiable services among multiple parties. Blockchain technology provides many desired features for large-scale IIoT infrastructures, such as decentralization, trustworthiness, trackability, and immutability. This paper presents a blockchain-based IIoT architecture to support immutable and verifiable services. However, when applying blockchain technology to the IIoT infrastructure, the required storage space posts a grant challenge to resource-constrained IIoT infrastructures. To address the storage issue, this paper proposes a hierarchical blockchain storage structure, \textit{ChainSplitter}.  Specially, the proposed architecture features a hierarchical storage structure where the majority of the blockchain is stored in the clouds, while the most recent blocks are stored in the overlay network of the individual IIoT networks. The proposed architecture seamlessly binds local IIoT networks, the blockchain overlay network, and the cloud infrastructure together through two connectors, the \textit{blockchain connector} and the \textit{cloud connector}, to construct the hierarchical blockchain storage. The blockchain connector in the overlay network builds blocks in blockchain from data generated in IIoT networks, and the cloud connector resolves the blockchain synchronization issues between the overlay network and the clouds. We also provide a case study to show the efficiency of the proposed hierarchical blockchain storage in a practical Industrial IoT case.
\end{abstract}

\IEEEpeerreviewmaketitle
\section{Introduction}
\label{Sec:Intro}
Internet of Things (IoT) is a paradigm that heterogeneous physical objects are interconnected through wired or wireless technologies, and further seamlessly connected to the Internet, enabling anywhere and anytime connectivity~\cite{xu2018blockchain}. In recent years, we have witnessed the wide adoption of IoT applications across a variety of industry sectors~\cite{da2014internet}, including manufacturing, home automation, transportation, and healthcare, to name a few. Most existing large-scale industrial IoT (IIoT) infrastructures nowadays are developed, deployed and maintained by individual parties. They are typically cloud-based and rely on centralized communication models~\cite{botta2016integration}, in which all devices are identified, authenticated, and connected through cloud servers that provide abundant computation and storage capacities. 

Along with the rapid growth of the size and complexity of IIoT networks, centralized IIoT solutions however are becoming more expensive due to the high deployment and maintenance cost associated with the network and cloud infrastructures~\cite{botta2016integration}. This problem is further exacerbated by the growing demand that IIoT networks from different parties should be able to communicate and collaboratively provide immutable and verifiable data. 
With traditional IIoT networks, data provided by individual industrial parties may not be trustworthy because they can be forged or modified by attackers or the owner of the data{~\cite{zhang2014iot}}. It is desirable to have mechanisms to verify the trustworthiness of data in IIoT networks.

Considering these requirements of IIoT networks, the emerging blockchain technology can serve as a promising candidate to provide immutable and verifiable services~\cite{swan2015blockchain}. Blockchain is a distributed data structure comprising a chain of blocks which is based on a decentralized peer-to-peer (P2P) network. It removes the need of a central controller, and also allows parties to transact even though they may not trust specific individuals. 
By introducing the blockchain technology into IIoT solutions, the management of numerous unspecified devices and processes -- including transactions and communications among the devices -- will become much easier.
Cryptocontract (e.g., self-executing, self-enforcing protocols) among IIoT devices can be recorded on blockchain as a smart contract~\cite{christidis2016blockchains} and executed automatically to greatly improve transaction efficiency~\cite{xu2018blockchain}. 

The blockchain technology, however, cannot be directly incorporated into existing IIoT solutions, given the extremely constrained resources in IIoT networks, and the requirement of the blockchain technology that each participant must keep an exact copy of the blockchain to guarantee the consistency. 
For example, Bitcoin is one of the most successful blockchain-based applications to date. The Bitcoin blockchain, even though not updated frequently, currently contains more than 190GB of data, of which only some are related to currency transfers.  
Compared with Bitcoin, most large-scale IIoT systems generate much larger data volumes, and it is simply infeasible to store all the blocks in local IIoT networks. 
It is thus important to remove layers of inefficiency from the traditional storage structure in blockchain and design a new method which can scale out. 

Motivated by these challenges, this paper presents a blockchain-based IIoT architecture, which uses the blockchain as a distributed ledger to maintain records of all transactions in the IIoT networks. The proposed architecture separates the IIoT infrastructure into three layers: local IIoT networks, the blockchain overlay network, and the cloud infrastructure.  To address the storage challenges in IIoT networks, a novel blockchain storage structure is proposed to store the blocks in a hierarchical manner: the majority of the blockchain is stored in the cloud to leverage its abundant storage capacity, while the most recent blocks are stored in the overlay network of the individual IIoT networks.
As the blocks continue to be appended to the blockchain, the percentage of each part is maintained dynamically, depending mainly on two factors: the size of the current blockchain and the size of the storage (e.g., disk) provided on consensus nodes.  
To seamlessly connect these three layers, the design details of two connectors, the blockchain connector and the cloud connector, are also presented. The blockchain connector in an overlay network prepares blocks from transactions (data generated in IIoT networks), and the cloud connector addresses the blockchain synchronization issues between the overlay network and the clouds. 

The rest of the paper is organized as follows. Section~\ref{Sec:IoTInfr} presents the existing cloud-based IIoT infrastructure. Section~\ref{Sec:Case} discusses the opportunities and challenges to integrate the blockchain technology into IIoT solutions. Section~\ref{Sec:Network} presents the proposed blockchain-based IIoT architecture and the proposed hierarchical blockchain storage structure. 
Section~\ref{Sec:BCConn} and Section~\ref{Sec:CloudConn} describe the functionalities of the blockchain connector and cloud connector, respectively.
Section~\ref{Sec:CaseS} provides a case study in one IIoT scenario to show the efficiency of the proposed framework. 
Section~\ref{Sec:Related} summarizes the related work. Section~\ref{Sec:Concl} concludes the paper and discusses future work.

\section{Cloud-based IIoT Infrastructure}
\label{Sec:IoTInfr}
Fig.~\ref{Fig:IoTCloud} gives an overview of a typical cloud-based IIoT infrastructure, which mainly consists of three layers: device layer, gateway layer and cloud service layer. The device layer comprises heterogeneous IIoT devices, varying from powerful computing units to extremely low-power microcontrollers. These devices are connected to the gateway layer through various wired and wireless networking technologies, such as ZigBee, BLE, Ethernet, etc. At the gateway layer, most companies and organizations deploy their own customized gateways to manage the local IIoT networks, aggregate the data, and serve as  the bridges to the clouds ~\cite{WebSmartThings}~\cite{WebWink}. 
These customized gateways are usually an integral part of the deployed IIoT infrastructure, which leads directly to ``stovepipe" solutions~\cite{johannesson2001design}. This further causes the interoperability issues, that is data and services provided by one organization cannot be shared or utilized by devices from the other organizations (due to different networking protocols, data formats, etc.), and the employed security mechanisms are often proprietary and undocumented.

For easy understanding and presentation, we use ``IoT" to represent ``IIoT" in the following description.

Traditionally, the device layer and gateway layer together form the local IoT networks. A typical local IoT network consists of the following four components (as shown in the left of Fig.~\ref{Fig:Netarch}): 

\begin{figure}
  \centering
  \includegraphics[width=8cm]{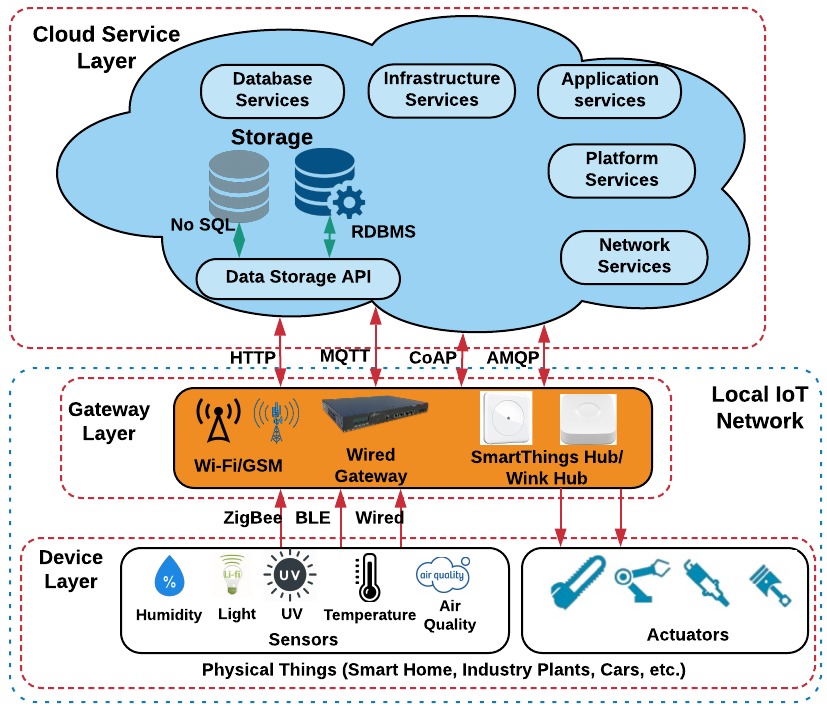}
  \caption{An overview of a typical cloud-based IIoT infrastructure.}
  \label{Fig:IoTCloud}
  \vspace{-0.1in}
\end{figure}

\noindent {\bf IoT Devices:}
\label{Sec:Device}
Most IoT devices are deployed in the physical world to measure and sample their associated physical or cyber objects. They have constrained resources, including memory size, computation power, and communication bandwidth~\cite{sehgal2012management}.
In addition, the devices and their adopted networking technologies are highly heterogeneous. This heterogeneity posts a grant challenge in interconnecting IIoT devices. It requires the interaction among the IoT devices to put the interoperability at the first place, such that heterogeneous devices are transformable in user's acceptable forms for both syntax and semantics~\cite{xiao2014user}.

\noindent {\bf IoT Storage:}
\label{Sec:Storage}
In a local IoT network, a centralized storage scheme is commonly adopted to manage the IoT data, instead of either local schemes (e.g., storing data within the local memory of IoT devices) or distributed schemes (e.g., storing data within some nodes with rich storage resources in the network). 
In a centralized storage scheme, the data are collected by the local gateway, and then sent to and stored in a local centralized storage.  In our scheme, the local centralized storage could be either a historian or a private data center, in which all data are stored locally and privately. { This centralized storage within a local IoT network can provide faster access to the recent data without accessing the cloud.} Where and how the local storage is deployed in the local IoT network depends on the system design specification.  

\noindent {\bf Data Engine:}
\label{Sec:DateEngine}
The data engine is a \textit{software} component that transforms incoming and outgoing raw data to and from the IoT devices into required forms. For example, in the proposed blockchain-based IoT architecture to be elaborated in Section~\ref{Sec:Network}, raw data are formed as transactions, and encrypted and uploaded to the clouds upon requests. The data engine can be deployed on the gateway or a stand-alone computing facility in the local IoT network. 
To guarantee the security in the local IoT network, the data engine also provides additional services, such as key management (e.g., distributing and updating keys to secure data transfer in local IoT network) and security mechanisms (e.g., authentication, authorization and audit services).

\noindent {\bf Gateway:}
\label{Sec:Gate}
In a typical cloud-based IoT infrastructure, the gateway is a connection entity that links the local IoT network to a cloud. In our proposed blockchain-based IoT architecture in section~\ref{Sec:Network}, the gateway plays two main roles. 
On one hand, it is the sink of the local IoT network, providing data management and network management functions; on the other hand, it also serves as a P2P node on the blockchain overlay network, providing proxy functions, such as routing information provisioning, node authentication, and multicast group management~\cite{ishikawa2005jupiter}. 

In addition to the device and gateway layers, the cloud service layer provides cloud-related functionalities, such as database service and application service,  to manage the data provided by the local IoT networks. Both the local IoT networks and the cloud service layer together comprise the most common existing cloud-based IoT infrastructure.
 
\section{Opportunities and Challenges to Integrate Blockchain into IoT}
\label{Sec:Case}

Blockchain, based on a decentralized P2P network and integrated with cryptographic processes, can offer many new features and improve existing functionalities of IoT systems. Since blockchain is built for decentralized environments, its security scheme is more scalable than traditional ones, and its strong protections against data tampering will help prevent rogue devices.  
The following features of the distributed architecture of blockchain make it an attractive technology for addressing many of the security and trust challenges in large-scale IoT systems~\cite{WebIBM}.

$\bullet$ Blockchain can be used to trace the measurements of IoT devices and prevent forging or modifying data.

$\bullet$ The IoT devices can exchange data through a blockchain to establish trust among themselves, instead of going through a third party. This significantly reduces the deployment and operation cost of IoT applications.  

$\bullet$ The distributed ledger structure of blockchain eliminates a single source of failure within the IoT ecosystem, protecting the IoT devices' data from tampering.

$\bullet$ Blockchain enables device autonomy via smart contract, individual identity, integrity of data, and supports P2P communication by removing technical bottlenecks and inefficiencies.

$\bullet$ Configuration of IoT devices can be complex, and the blockchain can be well adapted to provide IoT device identification, authentication and seamless secure data transfer.

Blockchain technology has enormous potential in creating trustless decentralized IoT applications, and provides lots of advantages from the technical perspectives. However, blockchain technology is still in its early stages, and there exist many barriers that limit the current blockchain technology from being applied to IoT applications. 

One of the most challenging problems is the storage issue when integrating the blockchain technology into IoT applications. 
In IoT scenarios, IoT devices can generate a huge amount of data in a short period, and both the data hash and the data itself need to be stored~\footnote{In most cryptocurrencies (e.g., Bitcoin), it is enough to only store the transaction hash.}.  
When a blockchain grows over time, all participating nodes will need larger storage capacity and higher bandwidth to keep up-to-date with the transactions added to the ledger, which is potential to become very costly. 
In this paper, we focus on solving the limited storage issue when introducing blockchain into the traditional cloud-based IoT architecture. In the following, we first use a \textit{heuristic} case study to illustrate the significance of the blockchain storage issue. 
Among many research topics emerging from the blockchain technology, Bitcoin is one of the most successful implementations of blockchain~\cite{bartoletti2017general}. A Bitcoin block consists of a block header with a size of 80 bytes and a list of transactions {as block payload (or block body)~\cite{velner2017smart}. And both block header and block body are included into the block.} Although the size of the block header is small, one of the major drawbacks of the existing Bitcoin witnessing scheme is that the auditors have to download the entire Bitcoin blockchain. As of November 2018, the Bitcoin blockchain contains more than 190GB of data, and it grows by 52GB every year~\cite{tomescu2017catena}~\cite{bcsize}~\cite{bartoletti2017general}.   
It is a challenging task if not impossible to download and store the whole blockchain in resource-constrained IoT gateways.  Thus, a more efficient blockchain storage structure is desired.

To better describe this challenge, we provide a numerical comparison, regarding the storage issue, between Bitcoin and a medium-size Industrial IoT (IIoT) system. In Bitcoin, the block size is currently limited at 1MB. The average size of a Bitcoin transaction, in one week of February 2019, is around 500 Bytes~\cite{WebTrade}.
Considering that the average number of transactions per block is 2000, and a Bitcoin block is generated by the miner around every 10 minutes, in every second it has 3.33 transactions generated within the Bitcoin network and thus the average data volume is 1.67KB per second, which is pretty mild. On the other hand, the comparing IIoT system is based on \textit{Emerson} Wireless Industrial Automation Systems~\cite{song2008wirelesshart}. 
We evaluate an industrial plant which has many wireless sensor and actuator networks (WSANs) deployed. We choose a medium-size system to estimate the average data volume, which consists of 50 WSANs, each having 100 nodes. We assume the average device sampling period is 1 second and the average message size is 100 bytes~\footnote{For simplicity, we only provide one \textit{heuristic} scenario.}. This leads to an average data volume of 500KB per second.   

{Assuming the block size is limited at 1 MB,} from the above comparison, we have the observation that, \textit{in one week}, the average {block} volume generated from Bitcoin network is about 1 GB, while the average block volume generated from the medium-size IIoT system is 302.4 GB, which is a huge amount of data that certainly cannot be stored in local IoT networks. It is worth noting that the required data for immutable and verifiable services are application dependent.  { Typically, these data  will be stored at least one year in the industrial case.}

In our proposed hierarchical storage structure for blockchain-based IoT systems, this issue can be addressed by a layered structure, where a blockchain overlay network is inserted in between the local IoT networks and the clouds. Resource-constrained IoT nodes only store a small portion of the blockchain that they need for their own transactions, while the majority of the blockchain is stored on the clouds. To ensure that this proposed layered structure work properly, interfaces (also called connectors in this paper) between the layers need to be carefully designed.  In the following, we first present the blockchain-based IoT architecture with hierarchical storage structure, and then describe the two designed connectors, the blockchain connector and the cloud connector.

\section{Blockchain-Based IoT Architecture}
\label{Sec:Network}
\begin{figure*}
  \centering
  \includegraphics[width=17.5cm]{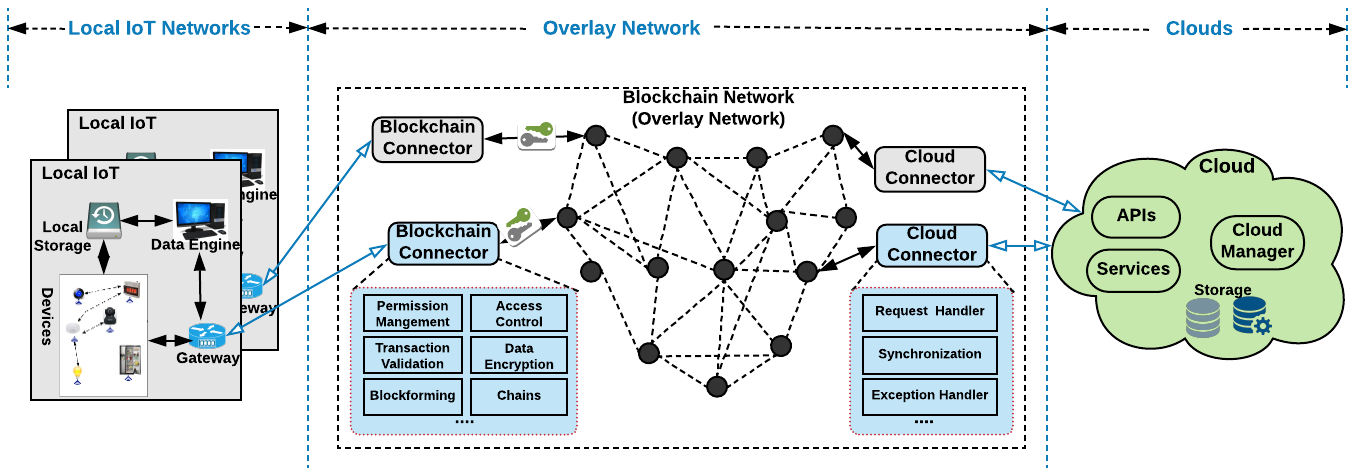}
  \caption{Overview of the blockchain-based IoT architecture. It consists of three components: local IoT networks, the blockchain overlay network, and clouds.}
  \label{Fig:Netarch}
  \vspace{-0.09in}
\end{figure*}

The proposed blockchain-based IoT architecture, as shown in Fig.~\ref{Fig:Netarch}, consists of three layers: local IoT networks, the P2P blockchain network, and the clouds. We first take a bottom-up approach to describe the key components of each layer, and then present the design details of the proposed hierarchical blockchain structure to mitigate the issue of storage capability.

\subsection{System Architecture}
\subsubsection{IoT Network}
In addition to their routine functions (see Section~\ref{Sec:IoTInfr}), the key added task of local IoT networks is to prepare transactions from the massive raw data collected from the network. Due to the heterogeneity of devices and applications in local IoT networks, the data need to be formatted and encrypted in a consistent format, so that the generated transactions can be easily operated and chained together in the blockchain overlay network. These operations are mainly done in the data engine. To seamlessly assemble the transactions into the blocks, the proposed architecture employs a \textit{blockchain connector} to serve as the interface between the IoT network and the overlay network (see Section~\ref{Sec:BCConn} for the details). 
{The \textit{blockchain connector} provides the functionalities to secure the transactions and the mechanisms to manage the devices access. Within the IoT network, this can be achieved by deploying a security platform to guarantee the authenticity of raw data from the authorized IoT devices. For example, in \textit{Emerson}'s blockchain project, it leverages the \textit{Mocana Security Platform}~\cite{WebMocana}, which is based on trusted hardware, to ensure the security of transactions.} 

Note that the proposed blockchain-based IoT architecture assumes that, {by utilizing the security platform,} securing and transferring the raw data into transactions happen in the local IoT network. The trust and security of local transactions can be guaranteed via security platform in a centralized and private manner. {And, the gateway, as a peer in the overlay network, sends the prepared transactions to the overlay network.} 
After obtaining the transactions, it is the responsibility of the overlay network (via the peers)  to create blocks, verify blocks, and chain the blocks together.  

\subsubsection{Overlay Network}
\label{Sec:Overlay}

After the transactions are published from the local IoT networks via the gateways, nodes in the P2P overlay network are responsible for forming the blockchain. The use of a P2P overlay network can provide several desirable features, such as scalability, high availability, and self-configuration~\cite{cirani2014scalable}.
The gateways of local IoT networks are organized as peers of the P2P overlay via \textit{Blockchain Connectors} (see Section~\ref{Sec:BCConn} for the details), which provide an efficient relay to construct the blocks from the transactions. The proposed architecture assumes that the overlay network follows the specified blockchain consensus protocols, e.g., PoW (Proof-of-Work) or PoS (Proof-of-Stake)~\cite{mingxiao2017review}, or Byzantine Fault Tolerance (BFT) schemes~\cite{castro1999practical}, depending on different IoT applications. 
Roughly speaking, BFT-based protocols do not have \textit{double spending} issues, which is more suitable for industrial IoT.

{Industrial IoT scenarios typically adopt a permissioned setting, which is operated by the known entities and imposes more stringent access control.  BFT-like consensuses have been widely investigated in permissioned blockchain~\cite{tseng2016recent} with the aim of outperforming PoW while ensuring adequate fault tolerance and faster finality of transactions. We assume the overlay network adopts a BFT protocol to achieve consensus process, such as scalable BFT for industrial secure metering~\cite{wang2019smchain}. And the threat model follows the definition in~\cite{wang2019smchain}, i.e., Byzantine nodes are less than $1/3$ of the total participating nodes.} 

\subsubsection{Clouds}
\label{Sec:Clouds}
By connecting the local IoT networks to the clouds, the IoT applications benefit from the virtually unlimited computing and storage resources of the cloud to compensate for its technological constraints (e.g., limited storage size, processing capability, and communication bandwidth). 
In our proposed architecture, the blockchain overlay network serves as the bridge that connects local IoT networks and the clouds together. To smoothly connect the overlay network to clouds, besides the basic functionalities of the cloud service layer (e.g., data storage, data management, etc.), the proposed architecture requires an additional bridging component, referred as \textit{cloud connector} to resolve the blockchain synchronization issues between the overlay network and clouds. 
The design details of the cloud connector will be presented in Section~\ref{Sec:CloudConn}.

{Notice that the clouds in our architecture are not maintained by a single group or entity.  Instead, the clouds are organized as a decentralized cloud storage, in which there not exists a single operator. To ensure the data consistency among the clouds, our scheme needs a decentralized object storage system to manage these clouds. \textit{Storj} network~\cite{storj} can serve as a promising candidate, which provides a robust object storage that encrypts, shards and distributes data to the distributed nodes for storage. \textit{Storj} storage management platform provides several key features, such as security and privacy, decentralization and Byzantine fault tolerance~\cite{storj}. These key features can meet the designated storage requirements.}

\subsection{Hierarchical Storage Structure of Blockchain}
\label{Sec:Arch}

As stated in Section~\ref{Sec:Case}, how to store the blockchain for large-scale IoT applications is a critical issue. In the original design of blockchain, every participating node has to store all the blocks locally. This storage mechanism is prohibitive when applying to large scale IoT applications, and a new  structure needs to be devised to address the  {storage} issue. {The main purpose of this paper is to handle the storage issues via hierarchical structure.} Before we present the proposed hierarchical storage structure for blockchain, we first make some assumptions on the overlay network. For example, assuming BFT algorithm as the underlying blockchain consensus protocol, the adopted BFT protocol should be robust enough to deal with a large number of participating nodes, and the ratio of Byzantine nodes is less than $1/3$.

To address the storage issue in the proposed blockchain-based IoT architecture, we adopt a hierarchical storage structure in which the blockchain data, or chain of blocks, are stored separately at two locations: the majority in the cloud storage and the latest portion in the peer nodes of the overlay network. Note that the local IoT networks might have the ability to store or download the blockchain, but it is not required in this proposed architecture. We assume that the main task of the local IoT network is to collect raw data generated from IoT devices and construct transactions. 

The constituent nodes in the overlay network, known as overlay nodes, can be either the IoT gateways or other {powerful} devices with much richer computing and storage resources, such as high-performance computers. The main task of the overlay nodes is to generate valid blocks that can be added into the blockchain. Different from the resource-constrained IoT devices, the overlay nodes have enough local storage to store a small portion of the blockchain, which is needed for computing new blocks and keeping them before sending them to the cloud.

\begin{figure*}
  \centering
  \includegraphics[width=16cm]{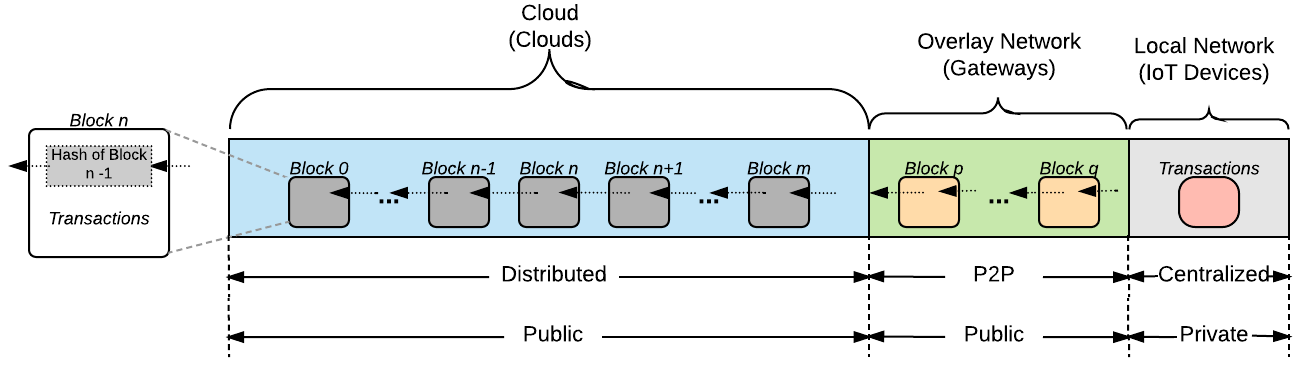}
  \vspace{-0.1in}
  \caption{Overview of the proposed blockchain structure.}
  \label{Fig:BCChain}
  \vspace{-0.1in}
\end{figure*}

To reduce the storage requirement, the overlay nodes only store the latest portion of the blockchain locally, and have the rest of the blocks in blockchain stored in the clouds. { Fig.~\ref{Fig:BCChain} shows an overview of the proposed blockchain structure.} It consists of three major parts: the first part is to deal with transactions in a centralized and private manner (e.g., information processing in gateways); the second part is to form the blocks according to the most recent blocks in a P2P overlay network (e.g., following consensus protocol); the last part is the majority of blocks stored in the cloud.  The authorized parties can access the data from the cloud and provide various smart services for IoT applications~\cite{dorri2017blockchain}. In our proposed hierarchical storage architecture, blocks can be uploaded and stored (or we can say synchronized) in multiple clouds while ensuring their consistency. The main purpose of using the cloud to store the blockchain data is to meet the storage capacity requirements from the large-scale IoT applications. Storing blocks in multiple clouds can prevent attacks where the data in a particular cloud (or a small set) are modified, if we assume most clouds are honest. {The proposed scheme uses \textit{Storj} storage, a distributed and decentralized cloud storage platform, to distribute and manage the data blocks in the clouds. Besides, other decentralized cloud storage platform can be used as a potential solution, such as IPFS file systems~\cite{benet2014ipfs}, to manage the data.} 

To seamlessly connect the hierarchical storage together, we need to define two interfaces (alternatively called connectors) to integrate these three layers together. 
The blockchain connector in an overlay network prepares blocks from transactions (data generated in IoT networks), and the cloud connector addresses the blockchain synchronization issues between the overlay network and the clouds.

\section{Design of the Blockchain Connector}
\label{Sec:BCConn}
The blockchain connector can be viewed as a middleware in the proposed architecture, residing on the overlay node. 
It has several essential functions to securely generate the blocks and chain them together. The functions include permission management, access control, transaction validation, data signature, and chain forming. 
In the following, we will describe these functions in detail. Note that the functionalities of the blockchain connector can be further extended to meet the requirements of different IoT applications. 

\subsection{Permission Management}

{Blockchain connector functions as an interface to the overlay network, and its permission management mainly serves to the local IoT network.}
In general, blockchains can be classified into permissioned and permissionless blockchains~\cite{cachin2016architecture}. 
In a permissioned blockchain, the permission management is required to manage the identity of its participants via certain procedures, such as whitelists or blacklists. 
In contrast, the participants in a permissionless blockchain are either pseudonymous or anonymous, like Bitcoin and Ethereum. However, using anonymous validators increases the risk of Sybil attack~\cite{levine2006survey}, where the attacker gains a disproportional amount of influence on the system. The protection mechanisms for Sybil attack, such as PoW consensus protocol, is costly and wasteful in permissionless blockchain. In general, the permissioned blockchain is able to legally host identity-related assets, while the permissionless system cannot.  Considering the unique features of most IoT applications, the proposed blockchain-based IoT architecture adopts the permissioned blockchain, which can efficiently deal with massive transactions, and only legitimate identities/nodes can construct blocks in the blockchain. {Also, most BFT consensus protocols require the permission management for the consensus nodes~\cite{castro1999practical}. }

Besides the permission of validation, other basic permissions, e.g., joining the network, submitting transactions, aggregating transactions into blocks, and creating assets, can also be managed by the permission management services. 

\subsection{Access Control}

In a permissioned blockchain, authentication and access control technologies are two critical mechanisms to address the security and privacy issues in IoT applications. 
 With access control, the users of the IoT applications will have full access to their data and have control over how the data will be shared. The user can assign a set of access permissions and designate who can query their blockchain. The access control permissions can be flexible and handle more than ``all-or-nothing" permissions~\cite{linn2016blockchain}. 
{Similar to permission management, the access control mainly targets to its local IoT network, and in a blockchain network, there no exists a centralized controller or authenticator.}

Many existing access control mechanisms can be adopted in the blockchain connector. For example, in~\cite{zyskind2015decentralizing}, the authors proposed an automated access-control manager using blockchain that does not require trust in a third party. The access-control manager ensures that users own and control their personal data and can perform the fine-grained access control, i.e., altering the set of permissions and revoking access to previously collected data. 
In~\cite{hardjono2016cloud}, the authors developed the \textit{ChainAchor} system which can address the issues of identity and access control within the shared permissioned blockchain. \textit{ChainAchor} can provide access control to entities seeking to submit transactions to the blockchain to read/verify transactions on the permissioned blockchain. 

\subsection{Transaction Validation and Data {Signature}}

The mechanism to validate transactions is specific to the blockchain. In general, the transactions are validated via being re-executed by the nodes that receive the blocks. For instance, the transaction validation mechanisms in Bitcoin rely on two general scripts: \textit{locking} and \textit{unlocking}~\cite{antonopoulos2014mastering}. Its verification scheme is mainly based on Unspent Transaction Output (UTXO) model, which defines an output of a blockchain transaction that has not been spent, i.e. used as an input in a new transaction.

Different from the complex transaction validation in Bitcoin, in IoT scenarios, the main task of the transaction validation is to prevent the data modification attacks in the P2P network. The transaction validation process generally can be divided into two phases: the initial verification and the transaction validation. The initial verification consists of validating the transaction's integrity by hashing the received transaction and comparing its hash against the hash value computed in the data engine. After the transaction passes the hash verification, the transaction validation process is simple; it adds a universal ``mark" and the new hash value to the policy header of the verified transaction. Since the verified transaction changes the mark field, it needs to compute the new hash of transaction. As long as the transaction validation finished, the transaction can be replayed to the neighbors. If the transaction fails to pass the validation process, the node in the overlay network might require the same transaction from the same gateway to perform the above validation process again. The validation scheme can set up an upward-boundary to the number of requests for the same transaction to avoid malicious gateways.

Due to the publicity of  {communication messages and the existence the malicious nodes}, it needs to deal with the {data's confidentiality integrity and authenticity (CIA)} when applying blockchain in IoT systems. The blockchain connector thus requires to {sign transactions (as a block) before sending out for verification}. The blockchain connector can be implemented on the local IoT gateway, thus  { the gateway must have the ability to perform the specified signature scheme}. This however needs the local IoT gateways to provide the cryptographic primitives to perform the  { digital signature, such as ECDSA scheme~\cite{johnson2001elliptic}}.

\subsection{Block and Chain Forming}

Another two important functions of the blockchain connector are the block forming and the chain forming.  The block forming is used to form blocks from { a set of} transactions while the chain forming is to chain the generated blocks together to form the blockchain.  

Once passing the transaction validation and the transactions are signed, the blockchain connector {in a gateway, as a consensus node in the overlay network,}  can propose blocks from the transactions according to specific rules or smart contract~\cite{kosba2016hawk}. 
{The verified and signed transactions are then broadcasted, 
via the blockchain connector, to the overlay network for consensus processes.}
Typically, devices in an IoT application might generate many transactions at the same time. { The consensus node (or gateway) should have enough buffer to store the incoming transactions from all consensus nodes.} The blockchain connector gathers the transactions that have been sent out over the blockchain network but have not yet been included in a confirmed block. {Depending on the specified consensus protocol,  the actual block and chain forming processes might be different. In the designated scheme, we adopt a BFT-based consensus protocol.}



{We briefly introduce how a BFT-based consensus protocol works for forming blockchain in our setting. BFT is an epoch-based consensus protocol, with the advantage of instant finality. In each epoch, one participating node (or gateway in our setting) is elected to be a leader and the leader coordinates each participating node. To simplify the description, a participating node and a gateway (equipped with blockchain connector) are the same entity as a consensus node in blockchain overlay network. In each epoch, the leader first verifies the integrity of the received transactions, and orders these validated transactions. A set of transactions are bundled together to form a block body, and the leader performs the Merkle tree hash on the block body to get a \textit{Merkle Root}. The \textit{Merkle Root},  together with other auxiliary information (e.g., the previous block hash, timestamp), are put into block header.  In each epoch, only one leader node can propose the block. Then, this proposed block is broadcasted to each participating node for verification and consensus process. Depending on the adopted BFT algorithm, it might have a different number of communication rounds among the participating nodes and leader. 
Typically, the blockchain connector can implement a smart contract to order these transactions, and the blockchain connector maintains the latest block he knows. Once the block is confirmed among the honest nodes, this block is added to the blockchain of each consensus node and the blockchain connector updates its latest block information for the next epoch block forming.

In case of the malicious leader, the consensus process goes into a \textit{view-change} procedure to elect a new leader. By leveraging BFT consensus, it can ensure the \textit{instant finality} of blockchain in each consensus round, which is more suitable for IoT scenarios.  By the property of instant finality, each consensus node keeps a same copy of blocks among all the honest participating nodes.
}
\section{Design of the Cloud Connector}
\label{Sec:CloudConn}
Similar to the blockchain connector, the proposed architecture also defines the \textit{cloud connector} between the overlay network and clouds, which updates and synchronizes the blocks into clouds. If we take the overlay network as a central element {or \textit{a black box}} in the proposed blockchain-based IoT architecture, the blockchain connector serves as an input source from local IoT networks, while the cloud connector serves as an output port to the cloud service layer. 

In this section, we focus on the introduction of the functionalities of the cloud connector. The cloud connector deals with the issues on when and how the blocks in the overlay network need to be synchronized to the cloud. In addition, it can provide mechanisms to deal with exceptions, such as the malicious overlay nodes in the overlay network. In general, the cloud connector includes three important components: the request handler, the synchronization engine, and an exception handler, as shown in Fig.~\ref{Fig:Netarch}. 

\subsection{Overview of the Cloud Connector}
\label{Sec:CloudConn1}

The deployment of a cloud connector is very flexible. It can be either deployed on the overlay nodes (like the blockchain connector), or integrated as a cloud management mechanism in the cloud service layer. 
Given that deploying the cloud connector in the cloud is subject to the single point failure attack in the proposed architecture, we install the cloud connector on the overlay nodes. 

The cloud connector is the interface between the overlay network and the clouds. It mainly includes a unified cloud protocol module for communicating with the overlay network. 
For secure communication between the overlay nodes and clouds, other mechanisms are still needed to guarantee the security. Here we only focus on when and how to synchronize the blocks in the overlay network to the clouds. 
From the connectivity perspective between the overlay network and clouds, it is not a good practice to perform synchronization immediately once a new block is generated. 
Instead, we intend to synchronize the blocks as a group based on the occurrence of certain triggering events or when some predefined time interval expires. Thus, our synchronization mechanism is mainly a hybrid scheme. Meanwhile, we assume that the overlay network always has a network connection with the cloud.  

To successfully synchronize the blocks between the overlay nodes and the cloud, the cloud connector needs to have the ability to handle when and how to perform the synchronization process. In the following subsections, the request handler is used to decide when the blocks in the overlay network need to be synchronized to the cloud; the synchronization engine is to deal with how the synchronization is performed; and the exception handler is to handle the potential exceptions during the synchronization process.

\subsection{Request Handler}

Different overlay nodes may have different storage for the latest blocks in the overlay network. In the case that there is not too much space for an overlay node to store new blocks, the node  broadcasts a message~\footnote{For example, this message indicates that its storage almost used up.}, {together with its latest block,} to the whole overlay network to notify a request to synchronize the blocks to the clouds.  {The peer nodes who receive this request check the received latest block with its own latest block. If both blocks are the same, the peer sends an ``agree to synchronization" response to the current leader. As long as the leader receives the same ``agree to synchronization" response from more than $2/3$ total peers~\footnote{Each communication message is signed by its sender.}. Then, the leader agrees to this synchronization request and broadcasts the aggregated response information (e.g., who vote to this synchronization). This process is actually a round of BFT consensus process. Note that each communicated message is signed by its sender, thus even a malicious leader cannot compromise the decision of synchronization from other nodes.} After the synchronization process among the overlay nodes, each overlay node has the same blockchain, which indicates that the last block in each overlay node is the same. To prevent the flooding attack by malicious nodes, we assume there exists a minimum storage capacity for individual overlay nodes so that a certain number of blocks can be stored.  It is recommended that each node in the overlay network has the same storage capacity; otherwise, there exists some resource wasted for the nodes with large capacity. According to this minimum storage capacity and speed of blockforming, a time interval can be set between two consecutive broadcasts. After the peer nodes receive the synchronization requests, they will connect to the cloud to synchronize the blocks. 

\subsection{Synchronization Process}
After the cloud receives a synchronization request from the overlay node, it starts the synchronization process. This process synchronizes the blockchain and the transactions at the same time. 

In the cloud connector, it has a field to specify the blockchain header, which is the most recently synchronized block in the cloud. In case that the cloud receives multiple synchronization requests, the cloud connector only needs to keep one copy of both the blockchain and its transactions. To resolve this issue, the basic idea is to let the cloud connector compare the blockchain header in the cloud with the current blockchain header from the overlay network. If both headers are the same, then the cloud is synchronized and denies this synchronization request; otherwise, the cloud will synchronize the blockchain and transactions from the overlay network. There exist several sophisticated works to solve file synchronization across multiple storages, such as MetaSync~\cite{han2015metasync} and UniDrive~\cite{tang2015unidrive}. 

Once the block synchronization process is finished, the cloud needs to send out the response message to notify the overlay network so that the overlay nodes{ know that the status of the current synchronization}. The response message typically has two types: regular response message and exception response message. The exception response message is used to notify the overlay network that errors happened during the synchronization process. This might require the repeat of the synchronization process. While the regular response message from the cloud needs to include in the most recently updated block information, the overlay nodes will only keep the most recently updated block, and swipe out all other blocks. We assume the communication between the overlay network and the cloud are secure {e.g., via secure communication channel or security platform.} Then, the overlay nodes consider this feedback block as the first block in its partial chain, and continue a new round block forming process until the storage space is full. {Due to the separation of the synchronization process and the consensus protocol in the overlay network, the overlay network does not stop the formation of the new blocks.}

\subsection{Exception Handler}
To guarantee that the blockchain is chained together as it was, the cloud needs to regularly perform a verification process to check if the currently synchronized blockchain is consistent. { This task is performed by the \textit{Storj} cloud platform within the clouds.} The exception handler is required to be performed before the cloud sends out the respond message { (as in the synchronization process)}. If some errors { (e.g., inconsistency among the clouds)} happen during the synchronization process, the cloud will send out the exception response message, otherwise, it will send out the regular response message. Compared to the nodes in the overlay network, the cloud typically has sufficient  computing resources. If an error is detected, the cloud will try other nodes to synchronize the blockchain and transactions. If the cloud detects  the error {(e.g., refusing to synchronization or synchronous wrong blockchain)} twice from the same overlay node, this node will be marked as a potential malicious node, and the cloud will notify the node administrators to inspect the malicious behavior.

\section{Case Study}
\label{Sec:CaseS}
This section provides an industrial case study to illustrate the efficiency of the proposed framework. We choose a medium-size industrial system, based on \textit{Emerson} Wireless Industrial Automation Systems,  
to estimate the average data volume generated when implementing the proposed hierarchical storage architecture~\footnote{We provide a quantitative analysis only regarding the storage requirements; other metrics, such as throughput and latency, are out the scope of this paper.}. The adopted system performs a continuous condition monitoring on industrial Smart Metering scenarios~\cite{8718376}, which consists of 50 WSANs (Wireless Sensor and Actuator Networks), each having 100 nodes. Each node is an industrial gateway, connected via SATA to a 2.5'' drive bay for storage.

We first present the industrial data structure for this case study (including transactions and blocks) and then provide the quantitative analysis on of the storage volume.

\subsection{Data Structure}
\label{Sec:DataS}

\subsubsection{Industrial Transactions}
\label{Sec:TX}

\Out{
\begin{figure}
  \centering
  \includegraphics[width=7cm]{Figs/TX_Str.png}
  \caption{Structure of a transaction.}
  \label{Fig:TX_STR}
  \vspace{-0.1in}
\end{figure}
}

Transactions in cryptocurrencies (e.g., UTXOs in Bitcoin~\cite{nakamoto2008bitcoin}) are quite different from industrial transactions, 
as they need to carry the industrial information on their own transactions. 
In the following description, we use a smart metering system as an example to outline the basic structure of an industrial transaction, which can be generalized into other industrial cases. Table~\ref{Tab:TX} shows a conceptual structure of the transaction with description.

\begin{table}[]
\scriptsize
\centering
\caption{Description of a Transaction}
\label{Tab:TX}
\begin{tabular}{|c|c|}
\hline
\textbf{Field}                                                             & \textbf{Description}                                                                                                                                                                                                                                                             \\ \hline
\textit{From}                                                     & \begin{tabular}[c]{@{}c@{}}The address of local metering device, \\ e.g., UUID of meters\end{tabular}                                                                                                                                                                   \\ \hline
\textit{To}                                                       & \begin{tabular}[c]{@{}c@{}}The target gateway, either field gateway\\ or edge gateway, that the metering\\ measurement is sent to\end{tabular}                                                                                                                          \\ \hline
\textit{Type}                                                     & What type of measurement, e.g, warning                                                                                                                                                                                                                                  \\ \hline
\textit{\begin{tabular}[c]{@{}c@{}}Device\\ \_info\end{tabular}}  & The information of metering device                                                                                                                                                                                                                                      \\ \hline
\textit{\begin{tabular}[c]{@{}c@{}}One\_Time\\ \_PK\end{tabular}} & \begin{tabular}[c]{@{}c@{}}The device's one-time public key used to\\ encrypt the message from device to gateway\\ so gateway can verify its integrity\\ and confidentiality\end{tabular}                                                                               \\ \hline
\textit{TimeStamp}                                                & \begin{tabular}[c]{@{}c@{}}Unix timestamp when a device is measured\\ its measurement (assuming all plants are) \\ synchronous locally. Also, a timestamp is\\ used to accept as valid if it is greater than\\ the timestamp from the previous data block\end{tabular} \\ \hline
\textit{TX\_ID}                                                   & \begin{tabular}[c]{@{}c@{}}To identify the order of measurement from\\ ``from" to the same ``to". Each measurement\\ has a unique ID during its block epoch\end{tabular}                                                                                                \\ \hline
\textit{Data}                                                     & Measured value from physical devices                                                                                                                                                                                                                                    \\ \hline
\textit{Hash\_Type}                                               & \begin{tabular}[c]{@{}c@{}}Indicate what digest algorithm used, e.g.,\\ SHA-256, SHA-512\end{tabular}                                                                                                                                                                   \\ \hline
\textit{TX\_Hash}                                                 & The digest of the measured value                                                                                                                                                                                                                                        \\ \hline
\textit{Sig\_Type}                                                & Indicate what signature algorithm used                                                                                                                                                                                                                                  \\ \hline
\textit{Signature}                                                & The signature of the measurement                                                                                                                                                                                                                                        \\ \hline
\end{tabular}
\vspace{-0.1in}
\end{table}

Here we specify some cryptography-related fields of Table~\ref{Tab:TX}. \textit{One\_Time\_PK} uses \textit{secp160r1} of Elliptic Curve Cryptography (ECC) (20 bytes); \textit{TX\_Hash} is based on SHA-256 algorithm (32 bytes); \textit{Signature} is  based on Boneh-Lynn-Shacham (BLS)~\cite{boneh2001short} signatures, which only require 33 bytes to achieve the same level security as 2048-bit RSA~\cite{gura2004comparing}.

\subsubsection{Industrial Block}
\label{Sec:Blocks}

A block in our industrial blockchain is called a ``data block". A data block is directly related to the transactions, which come from physical resources and local networks. Each data block consists of two parts: a block header and a block body. 
The header contains metadata about its block.  The body of the data block contains the transactions. These transactions are hashed only indirectly through the Merkle root. 
The description of each field of a data block is as shown in Table~\ref{Tab:DB}. 
Notice that most cryptocurrencies, e.g., Bitcoin, only store the transactions' hashes and the Merkle tree root into the blockchain, but industrial cases need the whole transaction to be stored in the data block for further analysis in condition monitoring.

\begin{table}[]
\scriptsize
\centering
\caption{Description of a Data Block}
\label{Tab:DB}
\begin{tabular}{|c|c|}
\hline
\textbf{Field}                                                            & \textbf{Description}                                                                                                                                                                                          \\ \hline
\multicolumn{2}{|c|}{\textit{Data Block Header}}                                                                                                                                                                                                                                          \\ \hline
\textit{\begin{tabular}[c]{@{}c@{}}Hash\_Pre\_\\ Data\_Blk\end{tabular}}  & \begin{tabular}[c]{@{}c@{}}Hash of previous data block. Each data blk\\ is inherited from its previous data block, \\ since it uses the previous block's hash to \\ create the new block's hash.\end{tabular} \\ \hline
\textit{Block Hash}                                                       & \begin{tabular}[c]{@{}c@{}}An identifier to identify a block,\\ which is a cryptographic hash.\end{tabular}                                                                                                   \\ \hline
\textit{Version}                                                          & \begin{tabular}[c]{@{}c@{}}The block version number, with which the\\ system can upgrade the software and\\ specify a new version.\end{tabular}                                                               \\ \hline
\textit{\begin{tabular}[c]{@{}c@{}}Merkle Root\\ of TXs\end{tabular}}     & \begin{tabular}[c]{@{}c@{}}Merkle tree root, a data structure that\\ summarizes the transactions in the block.\end{tabular}                                                                                   \\ \hline
\textit{\begin{tabular}[c]{@{}c@{}}No. of\\ TXs\end{tabular}}             & \begin{tabular}[c]{@{}c@{}}Identify the number of transactions to be\\ included in block body.\end{tabular}                                                                                                   \\ \hline
\textit{Signature}                                                        & \begin{tabular}[c]{@{}c@{}}The signature of the block, which is signed\\ by the creator of the block.\end{tabular}                                                                                            \\ \hline
\textit{Timestamp}                                                        & Show the time when a new block created.                                                                                                                                                                       \\ \hline
\multicolumn{2}{|c|}{\textit{Data Block Body}}                                                                                                                                                                                                                                            \\ \hline
\textit{No.}                                                              & \begin{tabular}[c]{@{}c@{}}Shows the order of transactions in one data\\ block sequentially from 1 to N, where\\ N is the total number of TXs in this block.\end{tabular}                                     \\ \hline
\textit{TX ID}                                                            & Extracted from Transaction.                                                                                                                                                                                   \\ \hline
\textit{TX Data}                                                          & Extracted from Transaction.                                                                                                                                                                                   \\ \hline
\textit{TX Hash}                                                          & Extracted from Transaction.                                                                                                                                                                                   \\ \hline
\end{tabular}
\vspace{-0.1in}
\end{table}

\subsection{Analysis}

\begin{figure}
  \centering
  \includegraphics[width=8.5cm]{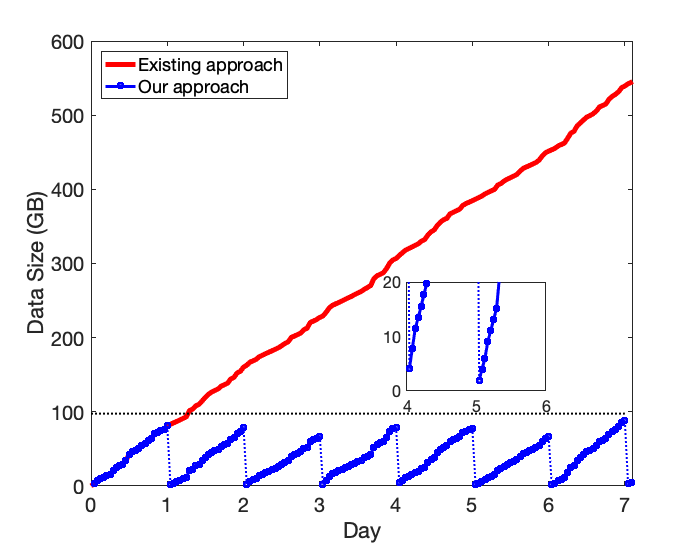}
  \caption{Data Storage Requirements for One-week's Data in a Medium-size Industrial Case.}
  \label{Fig:Exper}
  \vspace{-0.1in}
\end{figure}

A typical transaction size for smart metering in continuous condition monitoring is in the range of 120 $\sim$ 180 bytes for quantitative analysis. 
For example, if 150 bytes is the average transaction size and the average device sampling period is 1 second,  then the average data volume per second for a medium level industrial IoT platform is 
$50 \ WSANs * 100 \ nodes * 1 \ sample/second * 150 \  bytes = 750,000 \  bytes/second$ (750KB/second). 
Compared with the average data volume of 1.67 KB per second in the Bitcoin network, this storage requirement in the industrial case is pretty huge.

In our smart metering system, each consensus node (e.g., gateway) is equipped with one 128GB SATA Harddisk to store the blockchain. Considering the data volume generated and the size of the hard disk on each consensus node, the system updates the consensus node's data blocks to the cloud every 24 hours. Fig.~\ref{Fig:Exper} shows the block data storage requirement for one-week's evaluation in a medium-size industrial case on each consensus node. The existing approaches represent the generic solutions that store all the blockchain data on consensus nodes without resorting to storing the block data on external storage, such as cloud storage. 
As the figure shows, the blockchain storage requirement for existing approaches is huge. For a week period, the block data generated can be more than 500GB. Without effective backup storage, the old data will be overwritten by the newly generated blockchain, and one 128G disk is only large enough to store one day's data. Our approach utilizes the hierarchical blockchain storage to maintain the most recent blocks in consensus nodes, and the majority of the blockchain is synchronized into the cloud. We use the 200Mbps data link to synchronize the data from consensus nodes to the cloud. In this case study, it takes about 1 hour to upload one-day's blockchain to the cloud. As the synchronization processes and block forming processes in the overlay network happen simultaneously, the system does not need to stop. The subfigure in Fig.~\ref{Fig:Exper} shows the data volume change during the synchronization processes. We observed that the data volume cannot go down to zero during the synchronization processes, since the new blocks are still forming on the overlay network to link the blocks to the blockchain. 

\begin{figure}
  \centering
  \includegraphics[width=7.5cm]{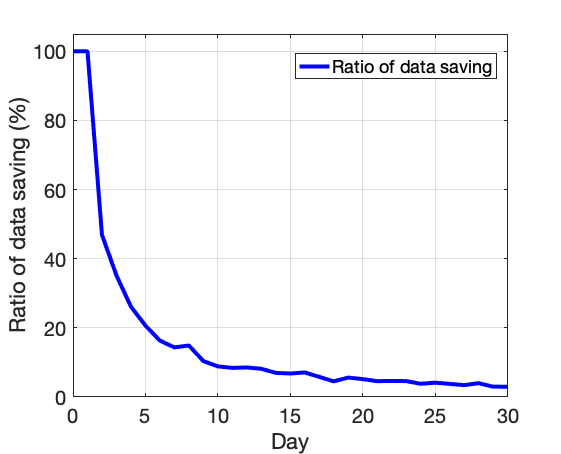}
  \caption{The Ratio of Data Saving Compared with the Existing Approaches.}
  \label{Fig:ratio}
  \vspace{-0.1in}
\end{figure}

Fig.~\ref{Fig:ratio} shows the overall percentage of the blockchain stored on consensus nodes as the time increasing over 30 days. From this figure, we can observe that the percentage of blockchain data stored on the consensus nodes is continually decreasing as time increases. Although the data volume of the whole blockchain continues growing, the blocks stored on the consensus nodes cannot be larger than a threshold because of the storage limitation of the consensus nodes. For instance, this threshold is 100GB in our case. When the blockchain data volume reaches this threshold, it triggers the synchronization processes to synchronize the recent blocks to the cloud. 
\section{Related Work}
\label{Sec:Related}
The blockchain technology was first introduced along with the Bitcoin by Satoshi Nakamoto~\cite{nakamoto2008bitcoin}. A tremendous amount of efforts from both academia and industry have contributed to the research of blockchain. When applying blockchain into IoT, it is still subject to several challenges, such as storage issues. This section provides the literature reviews on the blockchain storage. 

Some researches focus on increasing the usability by reducing the storage requirements on the \textit{client} side. In Bitcoin~\cite{nakamoto2008bitcoin}, Nakamoto proposes to use simplified payment verification (SPV) without running a full network node, and a user only needs to keep a copy of the block headers of the longest PoW chain. However, it does not mention how to reduce blockchain storage on the \textit{full network node}, and its verification is more vulnerable if the network is overpowered by an attacker. 
Xu et.al.~\cite{xu2018blockchain} propose a blockchain-based storage system, \textit{Sapphire}, for data analytics applications in the IoT. They present an OSD-based smart contract (OSC) as a transaction protocol, in which object storage devices employ embedded processors in the devices to process apart from storing data. 

Lunardi et.al.~\cite{lunardi2018distributed} propose a distributed access control on IoT ledger-based architecture.  
Due to limited storage on the gateway, the authors propose to parameterize and define the amount of information to be stored in the local IoT ledger, which only stores the new information in the block ledger. However, the authors did not mention the solution to store old information to the external storage, and the way to achieve that.  Sharma et.al.~\cite{sharma2018software} propose a blockchain-based distributed architecture with Software-Defined Networking (SDN) enabled controller fog nodes at the edge of the network. By leveraging a distributed fog node architecture that uses SDN and blockchain, the proposed model tries to bring computing resource to the edge of the IoT network. This structure reduces the access latency to large amounts of data in a secure manner, however, it does not deal with the blockchain storage issues.

Wilkinson et.al~\cite{wilkinson2014storj} propose a P2P cloud storage network, \textit{Storj}, allowing users to transfer and share data without reliance on a third-party data provider. The storage network periodically cryptographically checks the integrity and availability of the file. In \textit{Storj}, it uses MetaDisk~\cite{wilkinson2014metadisk}, a blockchain for decentralization metadata storage, to keep the consistency among the clouds. Similar to \textit{Storj}, \textit{Sia}~\cite{vorick2014sia} also is a platform for decentralized storage, which enables the formation of storage contracts between peers. In \textit{Sia}, it requires the storage providers to prove, at regular intervals, that they are still storing their client's data.
In our proposed scheme, we use \textit{Storj} to manage the cloud storage to maintain the consistency of blockchain among clouds.

Although a large amount of work has been contributed to integrating the blockchain technology to IoT systems and applications, to the best of our knowledge, none of them has dealt with the storage issue which is \textit{purely} based on the blockchain structure. Our work is the first attempt to mitigate this issue on the \textit{chain level}.
\section{Conclusion and Future Work}
\label{Sec:Concl}
Blockchain technology and the application of the technologies in IoT systems have gained a great deal of attentions from both academia and industry. However, it is a challenging problem to store and manage blockchain in IoT networks, due to the massive data generated from IoT applications and the limited resources in the IoT infrastructures. This paper proposed a hierarchical storage structure to store the majority of the blockchain in clouds, and maintain the most recently generated blocks in a blockchain overlay network. We further present a blockchain-based IoT architecture to maintain both blocks and transactions generated by the IoT networks. Two  software interfaces, a blockchain connector and a cloud connector, are defined to construct the blocks and synchronize them to the clouds. As a future work, we plan to  work on the implementation of the proposed blockchain-based IoT architecture in more real IoT applications and thoroughly evaluate its other performances, such as latency and throughput.

\bibliographystyle{IEEEtran}
\bibliography{refs}

\end{document}